\documentclass[]{aa}
\usepackage{txfonts}
\usepackage{natbib}
\usepackage{graphicx}

\begin{document}

\title{Dust and star formation in the centre  of NGC 3311
 \thanks{Based on observations taken at the European Southern Observatory, Cerro Paranal, Chile, under  programme 094.B-0711(A).
}}
    
\subtitle{}

\author{
T. Richtler,     \inst{1} 
M. Hilker,   \inst{2}
M. Arnaboldi \inst{2}
\and
C.E. Barbosa, \inst{3}
}

\offprints{T. Richtler}
\institute{
Departamento de Astronom\'{\i}a,
Universidad de Concepci\'on,
Concepci\'on, Chile;
tom@astro-udec.cl
\and
European Southern Observatory,
Karl-Schwarzschild-Str.2,
85748 Garching,
Germany
\and
Universidade de S\~ao Paulo, IAG, Departamento de Astronomia, Rua do Mat\~ao 1226, S\~ao Paulo-SP, Brazil
}

\date{Received  / Accepted }

\abstract
 { NGC 3311 is the central galaxy of the Hydra I galaxy cluster. It has a hot interstellar medium and hosts a central dust lane with emission lines.  These dust lanes are frequent in
 elliptical galaxies, but the case of NGC 3311 might be particularly interesting for problems of dust lifetime and the role of cool gas in the central parts.    }
  {We aim to use archival HST images  and MUSE data to investigate the central dust structure of NGC 3311. }
 {We used
the tool  PyParadise to model the stellar population  and extract the emission lines. }
 {The HST/ACS colour map reveals the known dust structures, but also blue spots, which are places of strong line emission.  A dusty 'mini-jet' emanates
 from the centre. The distribution of the emission line gas   matches the dust silhouette almost exactly. Close to the brightest H$\alpha$ emission,  the ratio $\rm [NII]/H\alpha$ resembles that of HII-regions; in the outer parts, [NII] gets stronger and is similar to  LINER{Low-ionization nuclear emission-line region} -like spectra. 
 
 The gas kinematics is consistent with that of a rotating disc.  The Doppler shifts  of the strongest line emissions, which   indicate the areas of highest star formation activity, smoothly fit into the disc symmetry.
  The metallicity is supersolar. The presence of  
 neutral gas is indicated by the fit residuals of the stellar NaI D absorption line, which we interpret as interstellar absorption.  We estimate the mass of the neutral gas to be of the order of the X-ray mass.    The dynamical mass infers a stellar  population of intermediate age, whose globular clusters have already been identified.   }
 { Our findings can be harmonised in a scenario in which the  star formation is triggered by the accretion of cold gas onto a pre-existing gas/dust  disc or ring. Newly produced dust then contributes to the longevity
 of the dust.
 }

\titlerunning{Dust and more in the centre of NGC 3311}

\maketitle

\section{Introduction}

 NGC 3311 is the  central galaxy of the Hydra I galaxy cluster.  Its  structural properties are naturally influenced by infall processes.
  For example, the
 rich globular cluster system  indicate the early dissolution of dwarf galaxies and actual infall processes are morphologically and kinematically visible
 \citep{ventimiglia10,ventimiglia11,misgeld11,richtler11,coccato11, arnaboldi12,barbosa16,barbosa18,hilker18}.   
 
 These previous works dealt with the large-scale structure of NGC 3311.
 However, the central parts are also extremely interesting. 
 NGC 3311 exhibits a central dust lane \citep{lindblad77,wirth80} with a major axis length of about 800 pc, which is aligned with the major axis of the galaxy's bulge. \citet{vasterberg91} performed the first CCD-photometry
and   spectroscopy of the central region of  NGC 3311. They found a blue light excess (which in the present paper appears as the 'blue spot') and ionised gas with a [NII]/H$\alpha$-ratio indicating  photons from massive
stars.   Moreover,  they used the velocity profile of the gas  as a rotation signature and quoted a  dust mass of $3\times10^4 M_\odot$. \citet{grillmair94} showed the first Hubble Space Telescope (HST) image of the dust structure and quoted 
$4.6\times10^4 M_\odot$ as a lower limit for the dust mass.
  \citet{ferrari99} estimated a relatively low dust mass
of $3100  M_\odot$,  while \citet{vandokkum95} quoted $1.4\times10^6 M_\odot$ for the sum of dust and gas.  There is also ionised gas, for which \citet{macchetto96} found the HII mass to be of the same order.

 This finding of nuclear star formation in a massive elliptical galaxy by \citet{vasterberg91}
uncovered a  phenomenon  that until recently was thought to be very rare. Meanwhile, many examples of star formation in central galaxies of galaxy clusters have been found (e.g. \citealt{fogarty15,donahue15,bonaventura17,runge18}).  In the local Universe,  NGC 3311 appears to be  the nearest example. 
Although cooling of the hot gas seems to be a self-evident process for providing cold gas for star formation, some authors  \citep{bonaventura17,runge18} argue against cooling flows as an immediate driver for central star formation, because the gas deposition rate by cooling is much higher than the observed star formation rate. 

The finding of intermediate-age globular clusters \citep{hempel05} indicates that star formation in NGC 3311 has been ongoing  for a long time, which poses interesting questions for the origin and nature of the dust.
 Dust is found frequently  in  early-type galaxies \citep{ebneter85,sadler85,ebneter88,goudfrooij94,vandokkum95, tran01, lauer05}.  The variety of dust content and morphological appearance is very large. It ranges from very striking galaxy-wide dust patterns 
 to tiny central dusty discs or rings. 
 The main dust producers are  conventionally thought to be winds from supernovae and from asymptotic giant branch (AGB) stars (e.g. \citealt{dellagli17}).
 These young populations are normally lacking in early-type galaxies, so the existence of dust  has been thought to testify to the previous infall of dusty material. A further complicating  circumstance is the very short
 timescale of dust grain sputtering by hot gas (typically 1-10 Myr), which would require recent dust injection for which there is no observational evidence.   
  \citet{hirashita15} and \citet{sansom19} amongst others provide a review  of the literature on dust in early-type galaxies.  
 However, recent work suggests the need to significantly change this picture.  \citet{michalowski19} measure  the dust removal time in early-type galaxies by considering  the age dependence of  the dust mass-to-stellar mass
 ratio. They arrive at a half-mass lifetime of 1.75 Gyr, so the existence of dust without young populations is normal (while  the physical processes of dust removal still remain obscure). 
 They conclude that the dust in their galaxy sample is likely to be of internal origin, but that the dust yields of AGB stars are not high enough to explain the dust masses, so additional processes like grain growth in
 the metal phase may be important.
  
The process of dust sputtering  is expected to be particularly efficient in NGC 3311 with its high central density of X-ray photons, but apparently this is not the case. 
The dusty centre  of NGC 3311 and its associated ionised gas deserves a  detailed investigation, especially  by capitalising on the resolution of HST images.  
 Data from the  Multi Unit Spectroscopic Explorer (MUSE) of the European Southern Observatory open the possibility  of studying the morphology of the dust and gas distribution, the gas metallicity, the ionisation sources, and the kinematical properties of the gas to infer its physics and origin.
To be consistent with our previous works, we adopt a distance to NGC 3311 of 50.7 Mpc, corresponding to a scale of 262 pc/arcsec. 

\section{Data and data reduction}
\label{sec:obs}
 MUSE is attached  to Unit 4 of the Very Large Telescope of the European Southern Observatory (ESO).   
MUSE is a mosaic of 24 integral field units (IFUs), covering a field of 1$\times$1 arcmin$^2$ in the wide field mode. The pixel scale is 0.2$\times$0.2 arcsec$^2$. 
 The spectral  resolution varies from R=2000 at 4700 \AA\ to R=4000 at 9300 \AA, which are the extremes of the spectral range covered.
The data consist of two exposures centred on NGC3311 with exposure times 1402s and 1418s, respectively. They were observed on the nights of 27 December 2014 and 15 January 2015. These data 
(programme 094.B-0711(A); PI: M. Arnaboldi) have already been used by \citet{barbosa18} and we refer the reader to this publication for more details.

The ESO Phase 3 concept offers pipeline reduced data products through the ESO science archive. One cube
(ADP.2017-03-27T12:49:43.627) combines the two exposures. However, the seeing is worse and because the signal to noise ratio (S/N) is
less important than the spatial resolution, we work with the data product  ADP.2016-06-20T22:49:26.325.fits (observed on the night of 27 December 2014), which shows a point spread function with a full width at half maximum (FWHM) =0.7\arcsec .
 The pipeline is described in the manual, version 1.6.2.  
 The basic reduction consists of applying {\it muse\_bias} and {\it  muse\_flat} (no correction for dark currents), followed by the wavelength calibration with {\it  muse\_wavecal}. 
 The line spread function has been calculated from the arc spectra using {\it muse\_lsf}. For the instrument geometry,  the tables are provided by ESO. Twilight exposures
 were used for the illumination correction, applying {\it muse\_twilight}.   The previous recipes produce frames/tables that  now enter the recipe {\it muse\_scibasic} that performs bias subtraction,
 flat field correction, wavelength calibration,  and more. The recipe {\it muse\_scipost} performs flux calibration and calculates the final data cube or as a choice, fully reduced pixtables that are combined by {\it muse\_exp\_combine} to
 produce a data cube with combined individual exposures. The pipeline also  corrects for  telluric absorption features.
We extract the spectra from the data cube using QFitsView.\footnote{www.mpe.mpg.de/~ott/dpuser/qfitsview.html.}

Moreover,  we use the following  data products from  the Hubble Legacy archive:  HST\_06554\_03\_wfpc2\_f814w\_pc\_drz.fits and HST\_06554\_03\_wfpc2\_f555w\_pc\_drz.fits  (proposal Nr: 6554;
PI: J. Brodie) as well as HST\_7820\_01\_NIC\_NIC2\_F160W (proposal Nr:7820; PI: D. Geisler).
These HST/planetary camera images cover the central
30\arcsec $\times$30\arcsec of NGC3311 with a pixel size of 0.05\arcsec. The exposure times are 3700 sec for the F555W-filter, 3800 sec for the F814W filter, and 10239.5 sec for the F160W-filter.
The data have already been used for work on globular clusters  in \citet{brodie00} and \citet{hempel05}.

\section{HST imaging}
\label{sec:hst}

\subsection{Structural properties of the dust lane}

 \begin{figure*}[t]
\begin{center}
\includegraphics[width=0.9\textwidth]{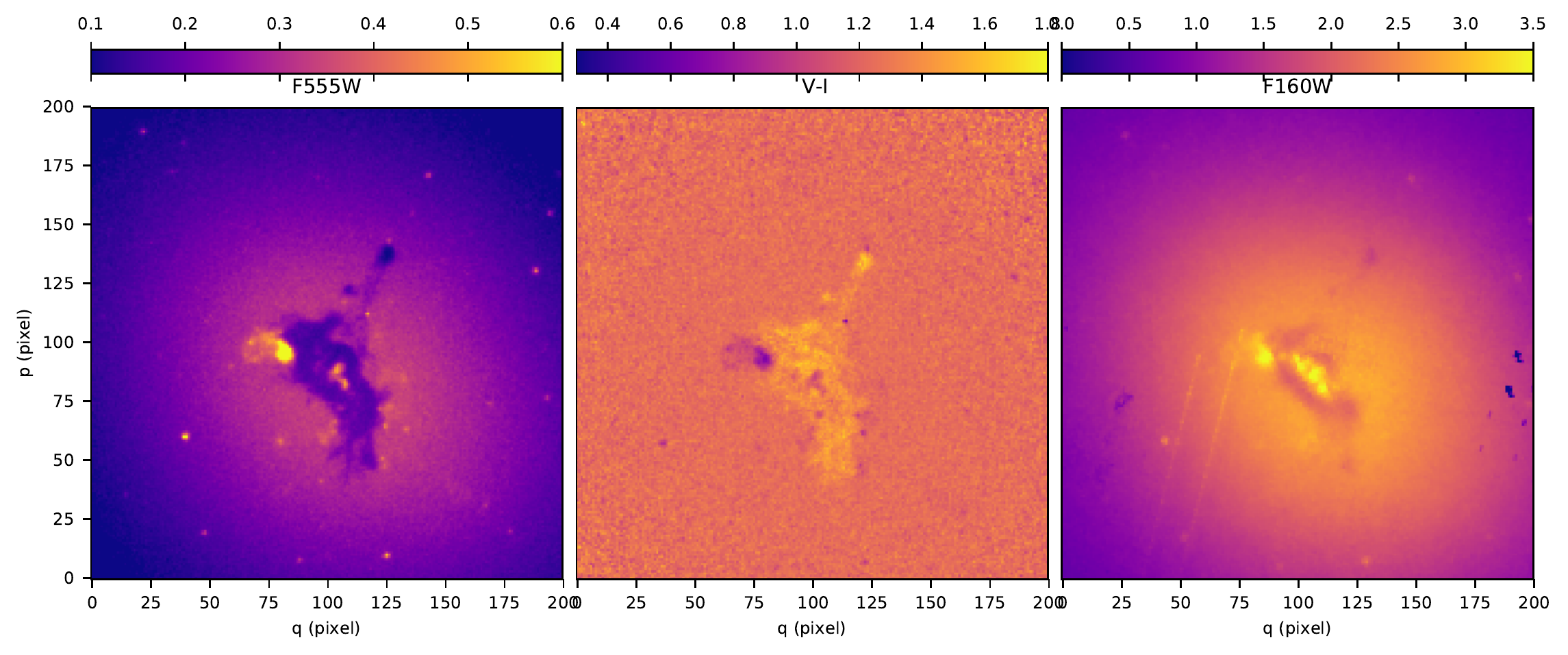}
\caption{Left panel: Central 10\arcsec$\times$10\arcsec (WFPC2, planetary camera, F555W) of NGC 3311. North is up, east  is to the left. The overall symmetry can be described as a disc with a radial feature pointing to the
north-north-west that seems to emerge  almost exactly from the centre. There is a population of about a dozen {\it bona fide} star clusters (little bright dots) fringing the border of the dust structure.
 Middle panel: V-I colour map. The blue spot is bluer by 0.5 mag than the  unreddened galaxy light. Right panel: H-band image of the same field as in the other panels. The blue spot is still the brightest object, 
 but three or four  other objects (globular clusters?) become visible.  Their integrated light appears as the "middle peak" in the paper. 
 }
\label{fig:hst}
\end{center}
\end{figure*}

The left panel of Fig. \ref{fig:hst} shows the central 10\arcsec$\times$10\arcsec\ section of the F555W-image (note that the same section has been shown previously by \citealt{grillmair94}, however, with a quality worse
than the present HST legacy product). The dust lane appears as a scraggy
elliptical structure. The extension of the semi-major axis is about 370 pc.  The position angle is $46.3^{\circ}\pm3^{\circ}$, the uncertainty being our estimate to define its major axis. This is in excellent
agreement with the    outer photometric isophotes of the stellar light, which is generally observed for central dusty discs. 
We employ the IRAF task {\it ellipse} to the F555W-image and get a smooth variation between a position angle of 52$^{\circ}$ at a radius of 50 pix (1310 pc)  and a position angle of 40$^{\circ}$ at
a radius of 200 pix (2600 pc).  The photometric centre of the isophotes within this radius interval coincides with an uncertainty of two to three pixels, with the optical brightness maximum near the centre of the dust lane.
A bright spot appears at the eastern border of the dust lane, which we will discuss in the next section.
 Two extensions of the central dust lane  
are striking: to the south a vertical structure, to 
the north-north west a straight and narrow dust band,  emerging from near the centre and ending in a knot-like feature with a total projected length of about 700 pc.
One also observes  a population of about a dozen {\it bona fide} star clusters, particularly in the southern part. \citet{brodie00} used the present images to investigate globular 
cluster metallicities, but do not comment on this very central population.

The middle panel shows a V-I colour map, constructed by -2.5$\times$ log(F555W/F814W) and transformed approximately to  V-I by adding 0.75 mag.  This matches the aperture photometry    
of \citet{prugniel98} who give V-I=1.33 for an aperture of 12\arcsec. 
The most striking feature in the colour map is the blueish colour of the bright spot, seen previously by \citet{vasterberg91}.
 It is bluer  than the
environmental stellar population by 0.5 mag.
The star clusters appear  typically a few 0.1 mag bluer than the galaxy light. 
This is in line with  \citet{hempel05}  who found a significantly high fraction of intermediate-age globular clusters in this
central region of NGC 3311.

The right panel shows the same field in the infrared H-band (F160W-filter).  One now sees that the blue spot is the least obscured one of a projected chain-like configuration of   four  
clump-like objects within the still visible dusty ring.  Their integrated light appears as the 'middle peak' in the paper.


\subsection{Bright blue spot in the optical}
This object appears in the Hubble Source catalogue (HSC) with the identifier ID 4201924376 and with W2-F555W=22.44 (ABMAG), which is almost identical to the V-magnitude.  
Because it is on top of the brightest regions in the galaxy,
its brightness and colour are subject to the exact definition of the background. To have an independent value, we applied a median filter of 10$\times$10 pixels to the planetary camera image
and subtracted this filtered image from the original. We performed aperture photometry with a radius of three pixels of selected point sources that are also HSC-objects to establish
a zero point.  Our result is W2-F555W=21.2,  significantly brighter than the HSC value. This corresponds to an absolute mag of $M_V = -12.3$  to which an extinction correction must still be applied.
We estimate the reddening  from the Balmer decrement in the extracted spectrum of the blue spot (Eq. \ref{eq:decre} and Table \ref{tab:ratios}). In ignorance of the reddening law of the dust, 
we identify the absorption in the F555W-filter with the absorption in the V-band and apply  Eq.\ref{eq:decre} to get $A_{F555W}$ = 1.4$\pm$0.4 mag, where the adopted uncertainty is
simply the uncertainty in the R-band from Table \ref{tab:ratios}. The real uncertainty may be larger.  The absolute magnitude is then $M_{F555W} = -13.7$ with an uncertainty that may exceed
0.5 mag  because of the additional distance uncertainty. Therefore the optical properties  are not well constrained and in the following we omit to give uncertainties.
The reddening is $E(V-I) = 1.6\times E(B-V)$ \citep{schultz75},   $V-I  \approx$ 0.8, and $(V-I)_0 \approx  0.06$. For comparison with recent stellar population models, 
we employ 
the webtool CMD (version 3.3)\citep{marigo17}. We use standard parameters, a Kroupa IMF,  UBVRIJHK as the photometric output , and the option of simulating a stellar population
of $10^4 M_\odot$. With  super solar abundance, as suggested by Fig. \ref{fig:diagnosis}, Z=0.04, and an age of $10^7$yr, the resulting colour is  $V-I = 0.55$. Lowering the age to $5\times10^6$yr,
the colour is $V-I = -0.24$. The respective mass-to-light ratios are $M/L_V = 0.05$ and $M/L_V = 0.026$.  The masses for the blue spot corresponding to  these models
 are $1.2\times10^6 M_\odot$  and $6.2\times10^5 M_\odot$. The light distribution is not symmetric. Ignoring this, a Gauss fit gives FWHM = 5.5 pixels. The point spread function is about two pixels, so five pixels or 65 pc is the approximate extension.  

We probably see an extended region of star formation where an  unresolved massive star cluster is embedded.  This may be a recent example of globular cluster formation in NGC 3311, which has been ongoing for many Gigayears. 
Within 1\arcsec\ to the west  of the blue spot, some faint loops/filaments  are discernible with one point source projected onto a filament. The colour appears not as blue as the blue spot, but distinctly bluer than
the galaxy light, which may be a population that originated in earlier star forming events.
 

\subsection{Stellar masses from infrared brightness}
In the H-band the blue spot is still the brightest object, but not longer extraordinarily bright. Assuming that there is no obscuration at all, one might derive a lower limit for the stellar mass. The field is too
crowded for precision photometry. We cover the blue spot with a window of 5$\times$5 pixels and the interior of the dusty ring with three adjacent windows of the same size. Subtracting the flux immediately outside the 
dust and following the instructions 
for getting photometry from NICMOS fluxes, we arrive at  $M_H = -14.2$ for the blue spot and $M_H = -15.2$ for the total brightness within the dust ring. 

 The above models that were used for optical photometry also provide the H-band.  The interpretational problem is now that the H-luminosity increases strongly within the age interval  $5\times10^6$ to
$10^7 yr$.  
  With the
above brightnesses, for the blue spot one gets  $1.6\times10^6 M_\odot$  for the $5\times10^6$ yr model and only  $5.5\times10^4 M_\odot$  for the $10^7$ yr model (we adopt a solar absolute H-magnitude
of 3.32). Corresponding to the difference in magnitude, the stellar masses inside
the dust ring for the two model ages are $4\times10^6 M_\odot$ and $1.4\times10^5 M_\odot$,
respectively.
It is also probable that older populations contribute in an unknown manner to the H-luminosity. 

\section{Morphology and distribution of ionised and neutral interstellar matter as seen with MUSE} 

\subsection{Separation of stellar light from interstellar emission or absorption }

To isolate the emission lines, the spectra of the galaxy light have to be fitted by a population synthesis approach and then subtracted. 
The interstellar NaI D absorption lines also  appear as residuals. 
 This can be done by various methods, for example STARLIGHT \citep{cid05} or PPXF \citep{cappellari17}, which are publicly available.
In this paper, we employ the PyParadise software \citep{husemann16}, which is an extended python version of Paradise. Details of the fitting procedure are explained in the appendix of  \citet{walcher15} to
which we refer the reader.

To prepare the cube for the spectral synthesis fit, its pixels are resampled by a Voronoi tessellation to produce areas of equal S/N. In our case the S/N was 15 and the number of bins was 30279.
The tessellation is visible in the Figs. \ref{fig:musemaps} and \ref{fig:EW}.


 \subsection{Strong emission lines}

 \begin{figure*}[t]
\begin{center}
\includegraphics[width=0.8\textwidth]{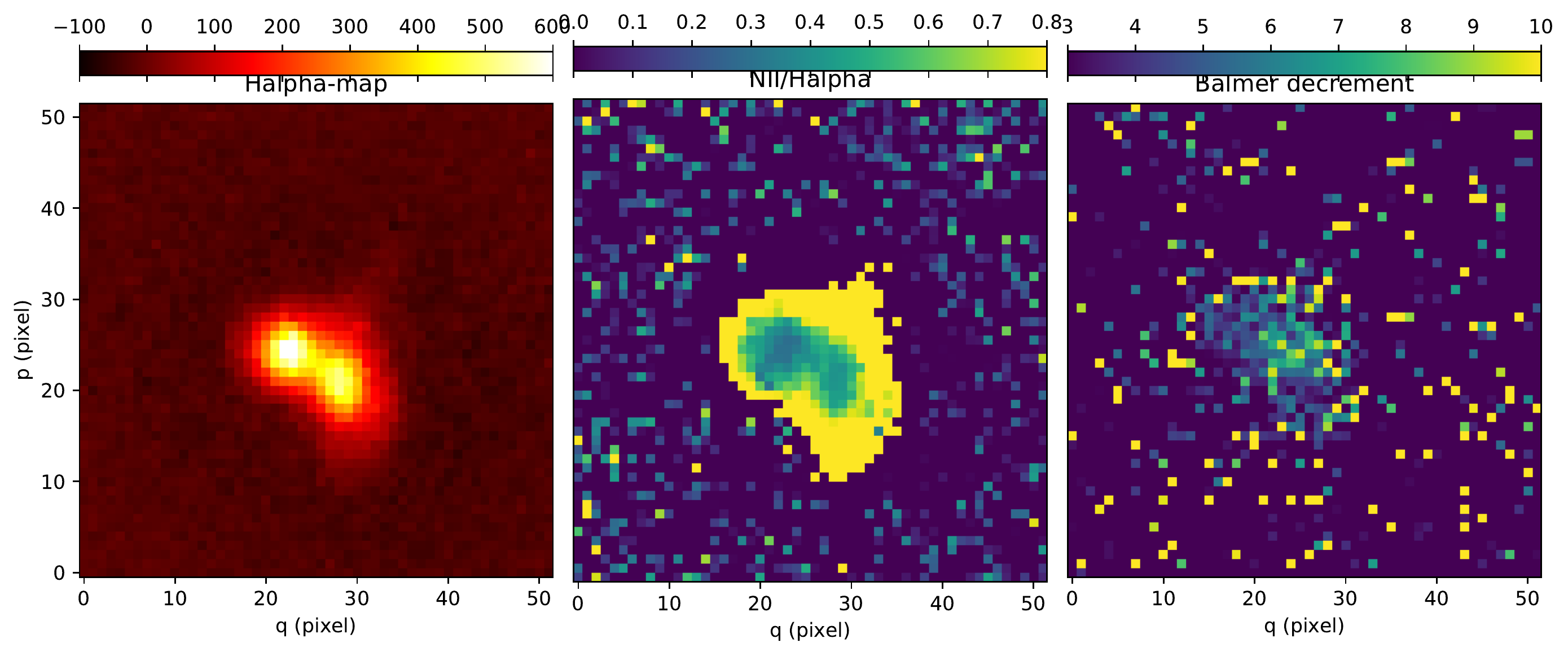}
\caption{Central 10\arcsec$\times$10\arcsec of the MUSE maps. North is up and east is to the left. Left panel: H$\alpha$-map. The dominant source is the blue spot, the second peak is approximately located near the centre of the dust lane. Middle panel: Ratio
[NII]/H$\alpha$. At the brightest H$\alpha$ peaks, the  ratio resembles an HII-region. In the outer parts,  [NII] gets stronger like in LINER-like spectra,  indicating the contribution of other
ionising sources like the post-AGB stars of the old stellar population. Right panel: Map of Balmer decrements mapping the extinction of the emission line region.  }
\label{fig:musemaps}
\end{center}
\end{figure*}

The most obvious features in the residual spectra are the strong emission lines, familiar from HII-regions: H$\beta$, [OIII]5007\AA, [NII]6549\AA, H$\alpha$, [NII]6583\AA, [SII]6617\AA, and [SII]6634\AA\ 
(see Fig. \ref{fig:residuals}).
Moreover, a residual of the stellar NaI D lines remains, which we interpret as an interstellar NaI D absorption (see next section).
 The most interesting maps are shown in Fig. \ref{fig:musemaps}. All emission features are strongly confined to the central region.

The main observation regarding the distribution of ionised gas is the close spatial coupling to the dust. This even includes the irregular extensions of a presumably central dusty disc
or ring-like structure. This finding is not special to NGC 3311, but is observed generally in early-type galaxies, as shown for example in  \citet{finkelman12}. 

In the brightest parts, the  emission line spectra resemble those of HII-regions. In the outer and fainter parts  of the dust structure, the low-ionisation lines get stronger, 
presumably due to a rising contribution from ionising sources like post-AGB stars and/or Low Mass X-Ray (LMXR)-binaries.

The spatial distribution of the ionised gas, on the other hand, is different from that of typical HII-regions where the dusty parts mark high density peaks in an overall more extended gaseous halo.
The close match of the gas extension with the dust extension speaks
against the origin of the gas being cooling out from the hot interstellar medium. 


\subsection{The NaI D  absorption lines}

\begin{figure}[h]
\begin{center}
\includegraphics[width=0.5\textwidth]{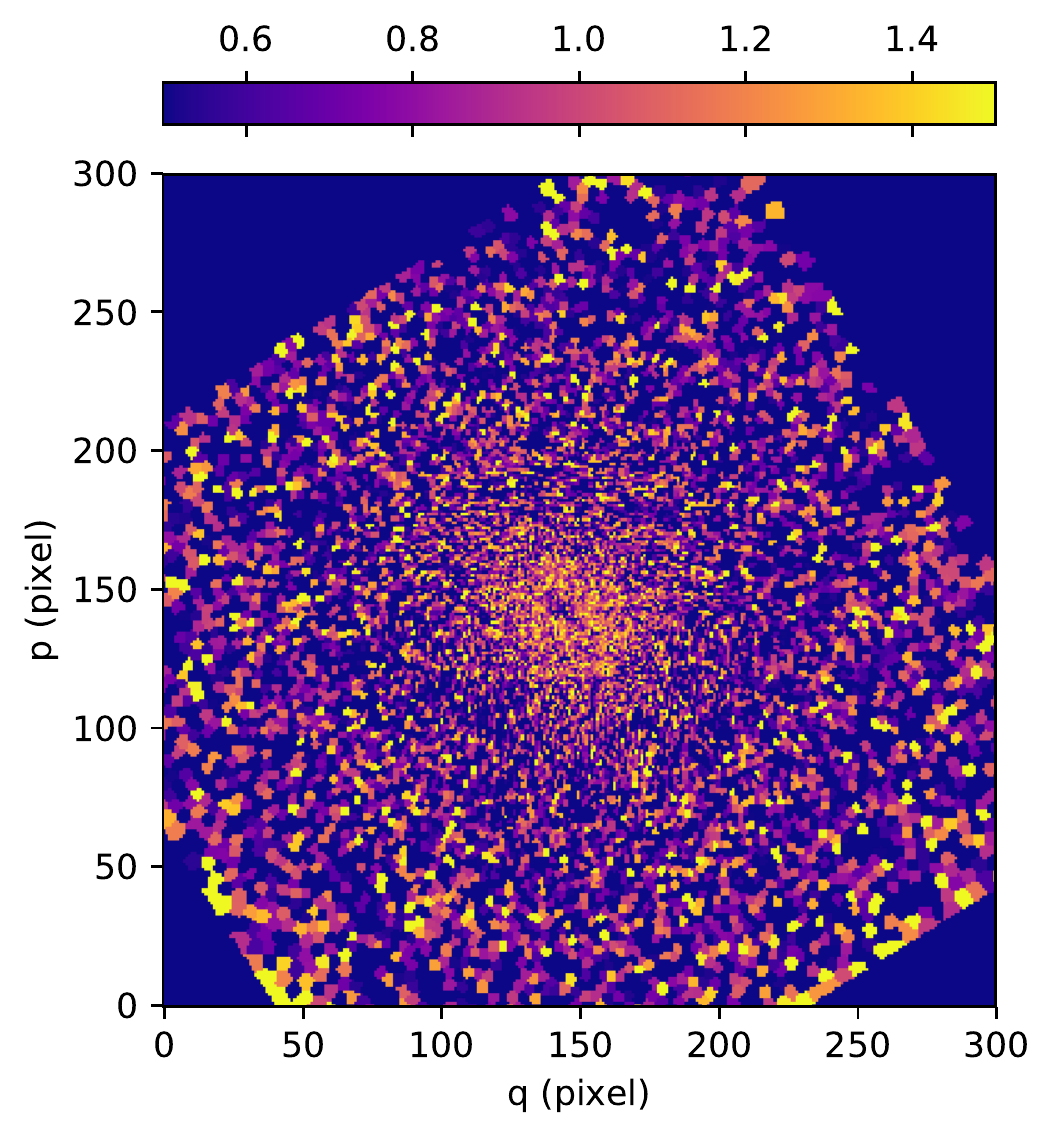} 
\caption{ A  1' $\times$1'  map of the equivalent width (EW, unit \AA) of the residual interstellar NaI absorption lines.  North is up, east is to the left, and the MUSE pixel scale is 0.2\arcsec. In contrast to Figs.1 and 2, we want
to show the largest area possible. The EWs refer to the sum of both lines of the doublet. Structures in the cold, atomic gas are not visible, except the 
 central dust lane that appears as a shadow. The lowest EWs are already about 0.5\AA, so we expect more cold gas outside the MUSE field. The gas probably has cooled down from the  hot X-ray gas. 
  }
\label{fig:EW}
\end{center}
\end{figure}

The NaI D doublet 5890\AA/5896\AA\  absorption is one of the strongest stellar lines. In each of the characteristic spectra (see Section \ref{sec:charac}), we observe that the stellar models do not fit the entire deep trough, but that   a strong broad residual remains. The reality of this
residual would mean the discovery of cold gas in a hot, X-ray bright environment  (see the discussion below for literature). While the occurrence of cold gas in the very centre may be plausible through the
short cooling times,   we also observe the NaI D residual as a quite extended feature. 
Figure \ref{fig:EW} shows a map of the equivalent widths of the NaI residuals. 
 This map has been constructed by  division by the model continuum and integration of the normalised flux over an interval of 40\AA.

Because the stellar line is so strong, one could suspect a  slight fitting defect of the stellar component  \citep{concas17}. 
An  argument for its reality is that in the region of the blue spot, the D1 and D2 lines get marginally resolved. The velocities of the component agree well with the NII velocity
of the same spectrum, which would not be expected if the residuum was artificial.
And finally, we have another MUSE example (NGC1316) where there is no doubt that the residual NaI D absorption traces atomic gas. More remarks are made on this in Section \ref{sec:cooling}.


\section{Characteristic spectra and line fluxes}
\label{sec:charac}
\begin{figure}[h]
\begin{center}
\includegraphics[width=0.5\textwidth]{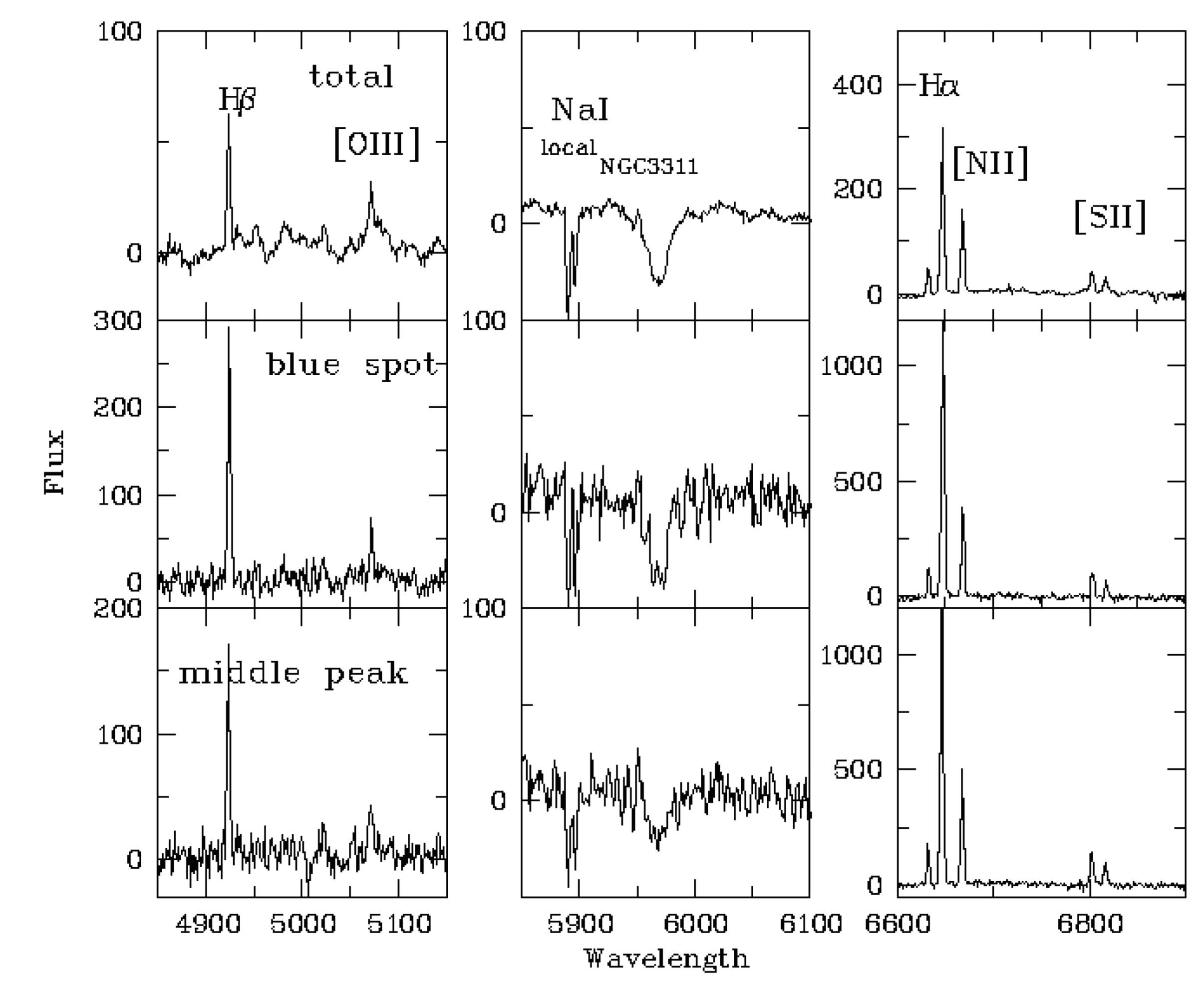} 
\caption{Residual spectra of three characteristic regions after subtracting the galaxy light are shown in those three spectral regions where emission lines are visible.  The appearance of the spectra resembles
that of a normal HII-region with H$\alpha$ being much stronger than [NII]. The fluxes are relative fluxes. 
  }
\label{fig:residuals}
\end{center}
\end{figure}

We define four characteristic spectra: a grand total with an extraction radius of 2.4\arcsec around the galaxy centre, the blue spot, the middle peak, and a spectrum with the blue spot and middle peak subtracted.
Figure \ref{fig:residuals} shows three of them.


\subsection{Line ratios and oxygen abundances}

Flux values are measured with {\it splot} under IRAF in the
continuum subtracted spectra. In order to avoid reddening corrections, we only consider line ratios constructed by neighbouring lines.
 Table \ref{tab:fluxes} lists the measurements. 
The comparison with photoionisation
models that predict strong line fluxes depending on the ionisation rate q\footnote{The ionisation rate has the unit number of ionising photons per unit area per second and per particle density.}
and oxygen abundance, is appropriate. \cite{dopita13} give a summary of the history of strong line diagnostics  and  a series of diagnostic graphs of which only a few are
interesting for us because of the restricted wavelength range. They recommend in particular the graphs [OIII]/H$\beta$ versus [NII/][SII] and [OIII]/[SII] versus [NII/[SII], which permit
the cleanest separation between  ionisation rate and  oxygen abundance. Here, the measurements refer to the sum of both [SII]-lines and the [NII]-line  at 6584\AA , respectively.
In addition we show the graph  [OIII]/H$\beta$ versus [NII]/$H\alpha$ to demonstrate that the line ratios are still in the range expected for HII-regions.

\begin{figure*}[th]
\begin{center}
\includegraphics[width=0.9\textwidth]{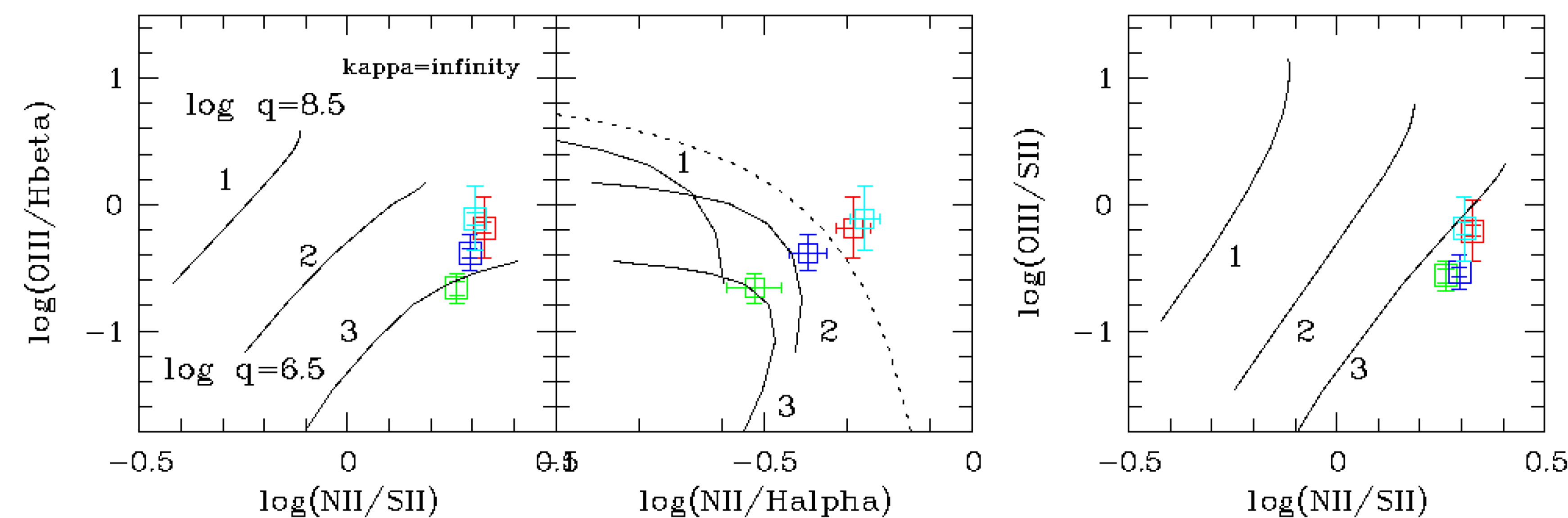} 
\caption{Diagnostic graphs for our four characteristic  spectra. Red: total, green: blue spot, blue: middle peak, cyan: the rest (see Table \ref{tab:fluxes}). The line ratios from Table \ref{tab:ratios} are compared
to the HII-region models from \citet{dopita13} (for a Boltzmann distribution of electrons) with the ionisation parameter log q and  three oxygen abundances in solar units (1x,2x,3x)  as parameters. The solid lines are lines of
constant abundance and varying log q. The range of
log q values is indicated in the left panel, which is  the same for all panels.   The dotted line in the middle panel is the line
separating HII-regions from AGN-like spectra according to \citet{kauffmann03}. 
 As  noted by  \citet{dopita13}, the right diagnostic graph seems to provide the cleanest separation 
of log q and oxygen abundance.  
Spectra 2 and 3 (blue spot and middle peak) show the characteristics of a  HII-region and a super solar metallicity in all graphs.  The other spectra  
 show contributions from other ionising sources, presumably post-AGB stars. The right panel is apparently insensitive to this circumstance. }
\label{fig:diagnosis}
\end{center}
\end{figure*}

A new feature in the models of \citet{dopita13}  is that the electron energies follow a $\kappa$-distribution, which in the case $\kappa=\infty$ is equal to a Boltzmann distribution. 
According to \citet{ferland16}, the thermalisation of electrons with a finite $\kappa$-value occurs  on shorter timescales than heating or cooling, which makes $\kappa=\infty$ the preferred choice.
For a Boltzmann-distribution of electron
energies, Fig. \ref{fig:diagnosis} shows  HII-region models of three different oxygen abundances (1,2,3 ; in solar units). The ionisation rates  range from log(q)=6.5 to log(q)=8.5.  The good separation of log(q) and oxygen abundance z in the left and right panel is obvious. Moreover,   both graphs appear to be insensitive to the LINER-contribution  because of the good agreement 
of abundances in spite of the large spread in [NII]/H$\alpha$. 

\begin{table*}[t!]
\resizebox{0.95\linewidth}{!}{
\begin{tabular}{ccccccccc}
\hline
\hline
ID & Spectrum  &  Extraction radius  & H$\beta$   & [OIII]  &H$\alpha$ &[NII] &    [S II] &  [S II]  \\
 & wavelength &   pixel &  4861      & 5007      & 6563 & 6583 & 6717 & 6731    \\
\hline
 1& total &  12 & 11.98 $\pm$  0.68& 7.81 $\pm$ 4.35&  51.92 $\pm$ 2.60& 26.88 $\pm$ 2.14& 7.88 $\pm$ 0.80& 4.76 $\pm$ 0.26 \\
 2& blue spot & 2& 1.24 $\pm$ 0.12& 0.27 $\pm$ 0.07& 6.04 $\pm$ 0.57& 1.81 $\pm$ 0.21& 0.65$\pm$ 0.05& 0.34 $\pm$ 0.01  \\
 3& middle peak &2 & 0.75 $\pm$ 0.02& 0.31 $\pm$ 0.10& 5.18 $\pm$ 0.36& 2.09 $\pm$ 0.15& 0.62 $\pm$ 0.03& 0.44 $\pm$ 0.01  \\
 4& the rest & - &9.41 $\pm$ 0.48& 7.29$\pm$ 4.26 & 41.29 $\pm$ 0.90&  22.82 $\pm$ 1.86& 6.44 $\pm$ 0.48& 4.83 $\pm$ 0.73  \\
\hline
\end{tabular}
}
\caption{Absolute line fluxes in units of 10$^{-16}$ erg/s/cm$^2$ for our four characteristic spectra. The flux values are the mean values  of two MUSE exposures. The uncertainties are the mean
square root values of the two deviations from the mean values. } 
\label{tab:fluxes}
\end{table*}

\begin{table*}[t!]
\resizebox{0.95\linewidth}{!}{
\begin{tabular}{ccccccc}
\hline
ID &   log ([OIII]/H$\beta$) & log([NII]/[SII]) & log[NII]/H$\alpha$)  & log([OIII]/[SII])  &H$\alpha$/H$\beta$ & $A_{H\alpha}$ \\
\hline
  1 &  -0.19 $\pm$0.24&  0.33 $\pm$0.05&  -0.29$\pm$ 0.04 &  -0.21  $\pm$0.24 & 4.33 $\pm$0.33 & 0.83 $\pm$0.46\\
      2 & -0.66 $\pm$0.12 & 0.26 $\pm$0.06 &  -0.52$\pm$ 0.06 & -0.56  $\pm$0.11 & 4.87 $\pm$0.66 &1.06 $\pm$0.41 \\
       3 & -0.38 $\pm$0.14 & 0.29 $\pm$0.03 & -0.39$\pm$ 0.04  & -0.53  $\pm$0.14 & 6.91 $\pm$0.51 &1.75 $\pm$0.28 \\
       4 & -0.11 $\pm$0.25 & 0.31 $\pm$0.05 & -0.26$\pm$ 0.04  & -0.19  $\pm$0.26 & 4.39 $\pm$0.24 & 0.85$\pm$0.46\\
\hline
\end{tabular}
}
\caption{Diagnostic line ratios  for our four characteristic spectra. The uncertainties  are calculated by error propagation from Table \ref{tab:fluxes}. The reddening values calculated from Eq. \ref{eq:decre} flux values are the mean values  of two MUSE exposures. The uncertainties are the mean
square root values of the two deviations from the mean values. } 
\label{tab:ratios}
\end{table*}

\subsection{H$\alpha$-flux}
\label{sec:halpha}
We calculate the total H$\alpha$  using  spectrum 1 in Table \ref{tab:fluxes}.
The observed  intensity within 12 pixels is $51.9\times10^{-16}$ erg/s/cm$^2$, corresponding to a flux of  1.5$\times 10^{39}$erg/s, in
very good agreement with \citet{macchetto96}, who give $51.9\times10^{-16}$ erg/s/cm$^2$.
This flux must be corrected for extinction. 
 We  estimate the reddening by using the Balmer decrements from Table \ref{tab:ratios}. The reddening is calculated adopting
the standard relation 

\begin{equation}
E(B-V) = 1.97 \times log ( ( H\alpha/H\beta)/2.85)  ) 
\label{eq:decre}
,\end{equation}


(e.g. \citet{momcheva13}).  Then we adopt 
  the reddening law of \citet{rieke85} (this may be incorrect, but there is no way of deriving the correct reddening law) and use the R-absorption as the absorption in H$\alpha$: $A_{H\alpha} = 2.3\times E(B-V)$.
  These values are listed as the last column in Table  \ref{tab:ratios}.  At our adopted distance, the total flux of spectrum 1 is $(3.3\pm0.7) \times10^{39} $ erg/s, where the distance uncertainty is estimated to be 10\%.
\citet{macchetto96} give an HII mass of 2100 $M_\odot $. However, to estimate the mass of HII from our data appears problematic. The normal procedure is to  use the H$\alpha$-flux, the electron density, and  the
electron temperature  to calculate the HII-mass, for example employing   Eq. (10) of \citet{kulkarni14} (see also \citealt{goudfrooij94,macchetto96}).  The electron density correlates with the ratio of the [SII] lines, 6717/6731.  \citet{proxauf14} give an updated relation (their Eq. 3). However, the line ratios in Table \ref{tab:fluxes} are too high and the corresponding electron densities  
much too low  to fit in the interval covered by the calibration relation.   One reason can be that  the subtraction on the synthesised galaxy spectrum leaves a residual  at the short-wavelength base of the 
S[II]6717-line. Stretching the uncertainties, a line ratio of  1.4 for the total spectrum is  a one-sigma deviation.   Adopting this value and applying the respective formulas of \citet{kulkarni14} and \citet{proxauf14},
one gets an HII-mass of $4.8\times10^5 M_\odot$ for  an electron temperature of 10000K, 
much higher than given by \citet{macchetto96}.

\subsection{HI mass}

We  estimate the mass of HI derived from the equivalent widths (EWs) of the residual NaI absorption. The residual nature without
any calibration makes it impossible to assess the robustness of such an estimate. 

For a relation between EW and hydrogen column density, we use
Eq. (1) of \citet{murga15} for the sum EW(D1+D2). With a reasonable precision (about 10\% for the interval 0.5 $<$ EW(D1+D2) $<$ 2.0), one can simply write
$$ N(HI)/[10^{21} cm^2] = 2\times EW(D1+D2). $$ 
\citet{murga15} emphasise that the relation already starts to saturate at EW=0.5\AA. Therefore we can only   estimate a lower limit. On the other hand,
the metal abundance is higher than in the Milky Way, which gives less HI per unit sodium. 
The mean EW in the  MUSE field (14.5$\times$14.5 kpc$^2$) is 0.96\AA. With the above formula this results in a mass of $3\times10^9 M_\odot$, which,  however is strongly concentrated towards the centre.
Whatever the exact number, the HI mass  within the MUSE field appears to be of the same order as
 the X-ray mass, which according to \citet{hayakawa06} is $2.2\times10^9 M_\odot$ within a radius of 14.5 kpc. 
 This is not much in comparison with the molecular gas mass found in cooling flows \citep{castignani20}. The fact that  the dynamical mass derived from X-rays using the assumption of hydrostatical equilibrium 
is lower than the baryonic mass alone \citep{richtler11}, now has an interpretation in that the X-ray gas alone does not provide hydrostatic equilibrium.
  More remarks on this are offered in Section \ref{sec:cooling}.



\section{Kinematics}

\begin{figure}[h]
\begin{center}
\includegraphics[width=0.5\textwidth]{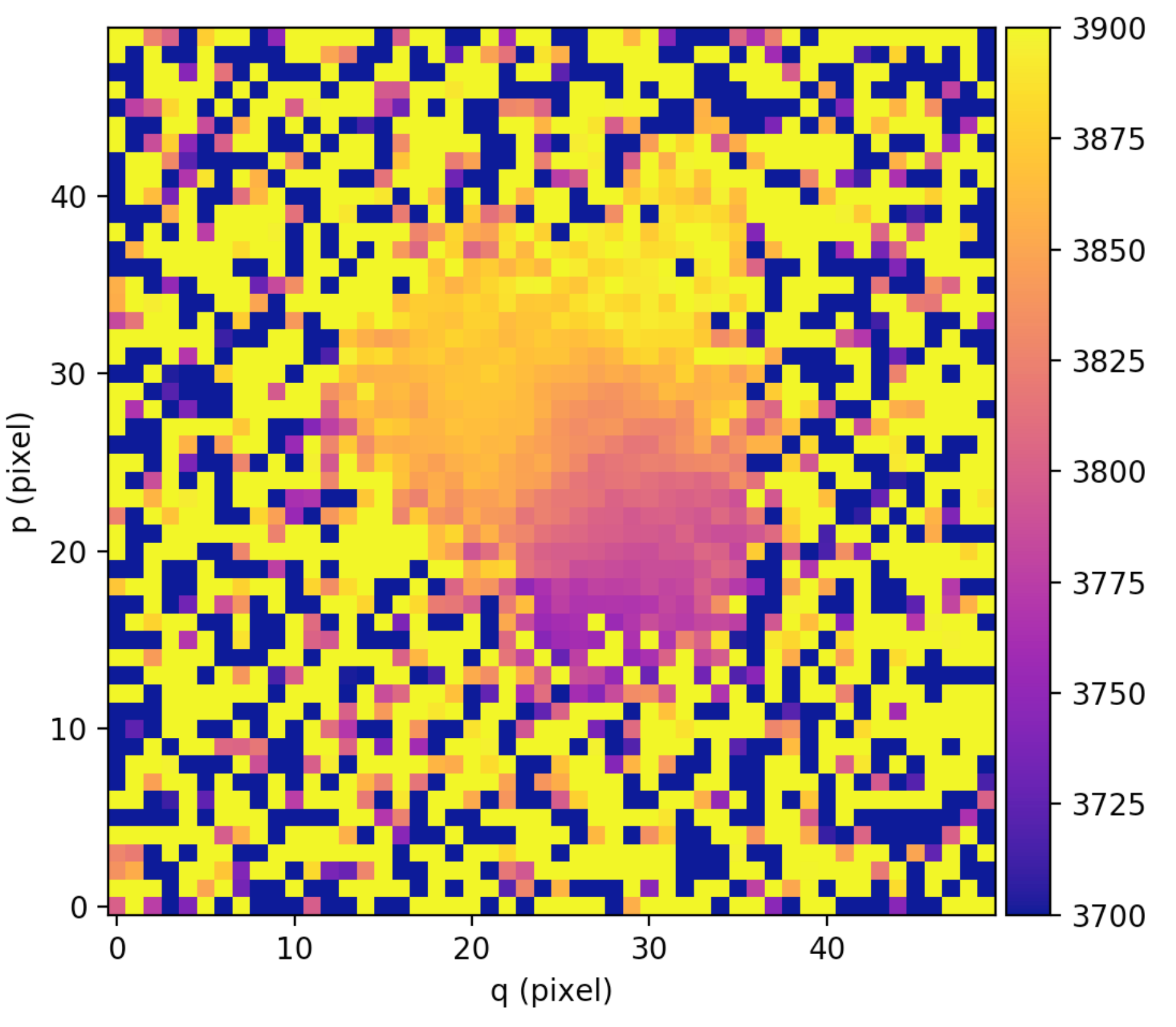} 
\caption{Velocity map in H$\alpha$. The size is 10\arcsec$\times$10\arcsec. North is up, east is to the left.   The appearance suggests  an inclined rotating disc, but the  derived circular velocity is too low for a good agreement with dynamical models for the stellar body. We note that the mini-jet 
has a higher velocity than the systemic velocity, which is consistent with  being expelled from the centre.
  }
\label{fig:velomap}
\end{center}
\end{figure}

We measure the velocity field   with  Qfitsview on the residual frame of PyParadise for H$\alpha$. Figure \ref{fig:velomap}  shows the velocity field. The first observation is that the kinematical symmetry resembles the  spatial  symmetry,
which gives strong support for the interpretation of a rotating disc.  Adopting $4.9\pm0.2\arcsec$ and $3.3\pm0.2\arcsec$ for the major and minor axis, respectively, the cosine of the inclination angle is 0.74$\pm$0.07.
 We measure  a projected rotation velocity of 45$\pm$10 km/s.  A  deprojection gives 61$\pm$15 km/s for the rotational velocity at a radial distance of 570 pc. The enclosed mass  is $M(r) = v_c^2 \times R/G$ with G = 0.0043
 with the units km/sec, years, and solar masses, which for R=570 pc gives  $4.93\times10^8 M_\odot$.  The model of \citet{richtler11} for the stellar light with a stellar $M/L_R$=6 and a distance of 48 Mpc encloses a mass of $1.87\times10^9 M_\odot$.
This is clearly not a good agreement, which would need a circular velocity of about 120 km/s. Possible explanations are an underestimated inclination angle and/or an intrinsic geometry that deviates from a circular disc. 
A further possibility is that the stellar M/L value is too high.
 A kinematical mass of $8.6\times10^8 M_\odot $ would still be consistent with our measurement (the  mass of the
supermassive black hole enters as a further unknown parameter, but see the remarks in Section \ref{sec:nucleardust}). This would require a mean $M/L_R$ value of around three in order
to agree with the photometric model. In other words, it demands a population mix equivalent to an  intermediate-age population of about 2.5 Gyr as a single stellar population (SSP). This is very interesting in view of the findings of \citet{hempel05}.
They studied the central globular cluster population of NGC 3311 using HST infrared imaging  (see Section \ref{sec:obs})  and identified a surprisingly large fraction of young- and intermediate-age globular clusters. This
makes the existence of a field population of an intermediate age highly plausible and fits our proposed scenario well (see section \ref{sec:scenario}).  

  The 'mini-jet' has a radial velocity of  3900 km/s,  higher than the velocities in the dust structure.  That is consistent with being expelled from the centre. If we assume the velocity relative
 to the systemic velocity to be 120 km/s, and the extension to be 2.8$\arcsec$,  the kinematic age (extension divided by velocity) is about $5\times10^6$ yr, under the assumption that the emission lines  trace the dust 
 kinematics.
 Also the southern extension 
 shows a velocity that smoothly connects with the disc.    

We also note that there is no striking interruption in the smooth velocity map that can be related to the blue spot, although its  light contribution is dominant. That indicates that star formation is an integral part of the rotating gaseous disc or ring.

\section{Discussion: A unifying scenario}

The old and new findings to be explained are: the mere existence of dust (old), alignment of the dusty disc or ring  with the bulge of the galaxy (old), supermetallicity (new), HII-regions from star formation (new), smooth HII-kinematics (new),  
the spatial match of dust and emission lines (new), and
the existence of
atomic gas (new). In the following, we discuss these findings in the light of the current literature.

\subsection{Cold gas in the hot halo of NGC 3311}
\label{sec:cooling}
The existence of cold atomic gas within a hot gaseous background is of course no surprise.  Cooling flows in the central parts of galaxy clusters  even show star formation at a much higher rate than is observed in
our case (e.g. \citealt{mcdonald15}). While the multi-phase nature of the hot  gas is expected \citep{gaspari13,gaspari17}, the interpretation of the residual 
NaI D absorption as being of an interstellar nature is not commonly accepted,  although the identifications of cold gas traced by sodium in M87   \citep{carter97} and NGC 4696  \citep{sparks97} are  robust.   \citet{jeong13} and \citet{park15} investigate 'sodium excess objects' in the Sloan Digital Sky Survey (SDSS) and conclude that
the sodium in early-type galaxies is indeed of a stellar nature in early-type galaxies, but of an interstellar nature in galaxies with active star formation. \citet{concas17}, on the other hand,
use the NaI D absorption as a tracer of cold gas kinematics in a large sample of SDSS galaxies. \citet{smith17}  identified  excess NaI D absorption in the central galaxy of the galaxy cluster Abell 3716.
 A further interesting case is NGC 6166, the central galaxy of Abell 2199. \citet{bender15}  noted the narrowness of the NaI D lines with respect to
other strong absorption lines. Without further analysis
they attributed that line to an interstellar medium.
 
 Cold  atomic gas in NGC 3311 is observed outside the central region, while the emission lines are strictly confined to the dust structure.  Star formation
needs molecular 
gas, but we do not know if molecules coexist like in other central galaxies  \citep{castignani20,olivares19}. Because molecules mainly form on the surface of dust grains (e.g. \citealt{cazaux04}),  it is plausible that 
 the dusty region provides the molecular gas needed for
star formation once the cold gas comes into contact with dust.

\subsection{Dust}
In the following, we discuss several aspects regarding the occurence of dust in NGC 3311.

 \subsubsection{Sputtering}
The mere existence of dust in the centre of NGC 3311 (and other  elliptical galaxies) would be difficult to understand if the sputtering timescale due to the
exposition of dust grains to a hot interstellar medium, as it exists in  NGC3311, were indeed extremely short.     Following \citet{hirashita17} (their Eq. 10) and adopting a central electron density of 11.7 \citep{hayakawa06}, a grain size of 1$\mu$m,  and a molecular weight of $\mu$ = 0.6, the sputtering timescale is  about 10$^4$ yr. At larger radii, the timescales are longer, the densities are lower, and in reality dust grains may be shielded, but sputtering of dust must be a  very rapid process, as has been assumed in recent simulations of intergalactic dust in galaxy clusters \citep{vogelsberger19}.
However, a dust removal timescale of 1.75 Gyr for early-type galaxies, as found empirically by \citet{michalowski19}, contradicts strongly the sputttering timescale, but the physical processes that control the dust removal
are still unknown and the origin of the dust is elusive.  
In the next sections, we will comment on different  possibilities. 
\subsubsection{Infall}
The alignment of the dusty disc with the major axis of the bulge in NGC 3311 (and many other cases)   argues against the dust being donated by an infalling dusty galaxy, because there is no physical reason for 
the angular momentum vector of the dust being parallel to the angular momentum of the much older stellar body.    Moreover, the close spatial match of ionised  gas and dust   suggests that the HII-gas  also contradicts this kind of scenario. 
  The low dust masses involved correspond to dwarf galaxy masses, while the high metallicity would demand a massive galaxy. As \citet{michalowski19} showed, dust can be old. It is therefore a plausible assumption that
  the present  dust  is the debris of a larger gaseous and dusty  structure that formed with the stellar body.   Central dusty discs are relatively frequent in elliptical galaxies without any  signs of a merger remnant. Such a scenario
  would predict the intermediate-age globular cluster population found by \citet{hempel05} to have a disc-like distribution, which could be easily tested.

\subsubsection{Dust from young stars}
This possibility demands that the dust has been produced by the  young stellar population through supernovae and/or AGB winds.  At these high metallicities, dust formation is very effective. Figure 2 of \citet{ginolfi17} shows that at even lower metallicities, a dust
mass of $10^3 M_\odot$  may be  produced by a stellar mass of $10^7 M_\odot$ (this dust mass does not respect a probably clumpy structure of the dust below the resolution limit, in which case the mass
can be much higher). This would be in at least partial agreement with a standard relation between $H\alpha$-fluxes and stellar mass. 
For the star formation rate, \citet{kewley02} give 
$$ SFR [M_\odot/y] = 4.5\times10^{-42} L(H\alpha)[erg/s], $$
which with our numbers translates into $1.7\times10^{-3} M_\odot/yr$. A stellar mass of  $10^7 M_\odot$ can  hardly be produced with this low star formation rate, but it could have been higher in the past. The probable existence
of intermediate-age populations \citep{hempel05} opens this possibility. The irregular shape of the dusty disc, on the other hand, may have been caused by stellar winds and supernovae from the young stellar population.



\subsubsection{Dust from nuclear winds}
\label{sec:nucleardust}
The dusty mini-jet  emanates from the nucleus. 
Dusty outflows from AGNs have been discussed for some time \citep{koenigl94}. In these models, accretion from the dusty torus is subject to radiative and hydrodynamic pressure and results in a bipolar
outflow. 
Modern parsec-scale IR interferometric observations of the active nuclei of nearby Seyfert galaxies   show the mid-IR emission coming mainly from the polar region of the 
supposed dusty torus (e.g. \citealt{hoenig13}). Outflow models with dusty winds have been presented that claim a much better agreement  with the morphology of highly resolved nuclear regions
of nearby Seyfert galaxies \citep{hoenig17}.


Could a nuclear dusty torus be the source for the dusty jet feature? NGC 3311 is too distant for a precise mass determination of its supermassive black hole (SMBH). We adopt $\sigma_c$ = 180 km/s as the central velocity dispersion (e.g.  \citealt{hilker18}) and  use Eq. (2) of \citet{graham11} 
to estimate the black hole mass, for which one gets $10^8 M_\odot$ with a scatter in dex of 0.37.  For the relation between the mass of the dusty torus and the SMBH mass, we refer to
Fig. 11 of  \citet{mor09},  where a loose correlation of the torus mass with the SMBH mass is visible.  A SMBH of $ 10^8 M_\odot $ corresponds to a torus mass of a few times $10^4 M_\odot$ of
dust. The dust mass of the jet feature therefore might easily be provided by a dusty torus. 


If the dust has its origin in the nucleus of NGC 3311, one should also expect to see similar features  in other galaxies. 
One can indeed easily verify by simply  looking at HST archival images that in many nearby dusty early-type galaxies, dust filaments emerge from the nuclear region. A particularly nice
example is NGC 5102. Surprisingly, there is hardly any literature on
this phenomenon.
A spectacular object in that respect is  NGC 1316 (Fornax A), where the NaI D lines precisely trace the dust structures (Richtler et al. 2020, in the revision process).        
 It is found not only in absorption, but also in emission, indicative of a galactic wind. There is much evidence that this wind is powered by a nuclear dusty outflow.




 




  
\subsection{Unifying scenario}
\label{sec:scenario}
Given the above points, the following scenario emerges. Cooling of the local hot gas provides a source  for material that becomes suitable for star formation only after  mixing
with the dusty disc, where dust cooling and molecule formation (e.g.  \citealt{cazaux04,vogelsberger19}) promote the formation of high density peaks (a variation of 'cold accretion').
Analogously to the cold accretion in galaxy formation models \citep{sancisi08,kleiner17}, such accretion could also alter the dynamics of a disc, being the cause of warps.   
 The star formation rate is controlled by the deposition rate  of dense, molecular gas,  which  might be lower than  the cooling rate of the X-ray gas.
Moreover, new young stellar populations provide dust and thus
contribute to the longevity of dust structures. 
The current star formation, which is evident from the HII emission line ratios,  was not a singular event  in the history of NGC 3311.  The existence of a stellar population component of intermediate age in the central regions of NGC 3311 is made manifest by its associated globular clusters \citep{hempel05}.


\section{Summary and conclusions}
On the basis of HST/ACS and NICMOS imaging as well as MUSE data cubes, we investigated the central region ($<$1 kpc) of NGC 3311, the central galaxy of the Hydra I cluster, focusing
on the central dust lane and emission lines. Our results are the following. The dust lane is a dusty ring-like feature that obscures an extended central star forming region. The brightest spot in the V-band appears strikingly blue on a  HST/ACS V-I colour map.  
 As the H-band image shows, it is the less obscured part of this region. 
 The stellar mass involved  in star formation  amounts to a few times $10^6$ solar masses.  
The  line emission  of the gas traces the shape of the dust pattern precisely.  
There is no ionised gas outside the dust features, which makes an external origin improbable. Therefore the dust must be old, which contradicts short sputtering timescales.  

The emission line ratios are typical for HII-regions, which is  proof that OB-stars are  the main ionisation sources. The oxygen abundances that are derived from strong line ratios are supersolar.   
The velocity  pattern of the gas indicates a rotating  disc. The rotation parameters demand a low stellar M/L, the SSP age  inside 500 pc being about 2.5 Gyr. This can be seen as a confirmation of
the existence of  an intermediate-age population that has previously been identified through globular cluster ages \citep{hempel05}. This moreover speaks for the longevity of the dust.
We observe a dusty mini-jet with a projected length  of    730 pc and a kinematical age of $5\times10^6$, which is kinematically distinct from the dust pattern and emanates
  from the centre.
  
Residual  interstellar NaI D-absorption is visible in the entire MUSE field, although strongly concentrated in the centre. This is one of the very few cases where cold gas in early-type galaxies 
becomes manifest through the sodium lines.  

All these findings are best understood within a scenario that describes the cooling  of hot X-ray gas in the inner halo that mixes with the dusty disc, experiences further cooling, and molecule 
formation. The subsequent existing young stars  provide more dust,  which contributes  to the longevity of the dust patterns.  



 \begin{acknowledgements}
We thank the anonymous referee for a constructive and helpful report. 
Bernd Husemann provided PyParadise. His assistance was essential for this work. 
We further thank the language editor for her careful work.
TR acknowledges support from  the BASAL Centro de
 Astrof\'{\i}sica y Tecnologias
Afines (CATA) PFB-06/2007.  This work has strongly benefitted from TR's visit at ESO/Garching under the  ESO science visitor programme.
TR cordially thanks the Astronomisches Institut der Ruhr-Universit\"at Bochum for hospitality. 
 \end{acknowledgements}

\bibliographystyle{aa}
\bibliography{MUSE.bib} 

\begin{thebibliography}{73}
\expandafter\ifx\csname natexlab\endcsname\relax\def\natexlab#1{#1}\fi

\bibitem[{{Arnaboldi} {et~al.}(2012){Arnaboldi}, {Ventimiglia}, {Iodice},
  {Gerhard}, \& {Coccato}}]{arnaboldi12}
{Arnaboldi}, M., {Ventimiglia}, G., {Iodice}, E., {Gerhard}, O., \& {Coccato},
  L. 2012, \aap, 545, A37

\bibitem[{{Barbosa} {et~al.}(2018){Barbosa}, {Arnaboldi}, {Coccato}, {Gerhard},
  {Mendes de Oliveira}, {Hilker}, \& {Richtler}}]{barbosa18}
{Barbosa}, C.~E., {Arnaboldi}, M., {Coccato}, L., {et~al.} 2018, \aap, 609, A78

\bibitem[{{Barbosa} {et~al.}(2016){Barbosa}, {Arnaboldi}, {Coccato}, {Hilker},
  {Mendes de Oliveira}, \& {Richtler}}]{barbosa16}
{Barbosa}, C.~E., {Arnaboldi}, M., {Coccato}, L., {et~al.} 2016, \aap, 589,
  A139

\bibitem[{{Bender} {et~al.}(2015){Bender}, {Kormendy}, {Cornell}, \&
  {Fisher}}]{bender15}
{Bender}, R., {Kormendy}, J., {Cornell}, M.~E., \& {Fisher}, D.~B. 2015, \apj,
  807, 56

\bibitem[{{Bonaventura} {et~al.}(2017){Bonaventura}, {Webb}, {Muzzin}, {Noble},
  {Lidman}, {Wilson}, {Yee}, {Geach}, {Hezaveh}, {Shupe}, \&
  {Surace}}]{bonaventura17}
{Bonaventura}, N.~R., {Webb}, T.~M.~A., {Muzzin}, A., {et~al.} 2017, \mnras,
  469, 1259

\bibitem[{{Brodie} {et~al.}(2000){Brodie}, {Larsen}, \&
  {Kissler-Patig}}]{brodie00}
{Brodie}, J.~P., {Larsen}, S.~S., \& {Kissler-Patig}, M. 2000, \apjl, 543, L19

\bibitem[{{Cappellari}(2017)}]{cappellari17}
{Cappellari}, M. 2017, \mnras, 466, 798

\bibitem[{{Carter} {et~al.}(1997){Carter}, {Johnstone}, \& {Fabian}}]{carter97}
{Carter}, D., {Johnstone}, R.~M., \& {Fabian}, A.~C. 1997, \mnras, 285, L20

\bibitem[{{Castignani} {et~al.}(2020){Castignani}, {Combes}, {Salom{\'e}}, \&
  {Freundlich}}]{castignani20}
{Castignani}, G., {Combes}, F., {Salom{\'e}}, P., \& {Freundlich}, J. 2020,
  \aap, 635, A32

\bibitem[{{Cazaux} \& {Tielens}(2004)}]{cazaux04}
{Cazaux}, S. \& {Tielens}, A.~G.~G.~M. 2004, \apj, 604, 222

\bibitem[{{Cid Fernandes} {et~al.}(2005){Cid Fernandes}, {Mateus}, {Sodr{\'e}},
  {Stasi{\'n}ska}, \& {Gomes}}]{cid05}
{Cid Fernandes}, R., {Mateus}, A., {Sodr{\'e}}, L., {Stasi{\'n}ska}, G., \&
  {Gomes}, J.~M. 2005, \mnras, 358, 363

\bibitem[{{Coccato} {et~al.}(2011){Coccato}, {Gerhard}, {Arnaboldi}, \&
  {Ventimiglia}}]{coccato11}
{Coccato}, L., {Gerhard}, O., {Arnaboldi}, M., \& {Ventimiglia}, G. 2011, \aap,
  533, A138

\bibitem[{{Concas} {et~al.}(2017){Concas}, {Popesso}, {Brusa}, {Mainieri}, \&
  {Thomas}}]{concas17}
{Concas}, A., {Popesso}, P., {Brusa}, M., {Mainieri}, V., \& {Thomas}, D. 2017,
  ArXiv e-prints

\bibitem[{{Dell'Agli} {et~al.}(2017){Dell'Agli}, {Garc{\'{\i}}a-Hern{\'a}ndez},
  {Schneider}, {Ventura}, {La Franca}, {Valiante}, {Marini}, \& {Di
  Criscienzo}}]{dellagli17}
{Dell'Agli}, F., {Garc{\'{\i}}a-Hern{\'a}ndez}, D.~A., {Schneider}, R.,
  {et~al.} 2017, \mnras, 467, 4431

\bibitem[{{Donahue} {et~al.}(2015){Donahue}, {Connor}, {Fogarty}, {Li}, {Voit},
  {Postman}, {Koekemoer}, {Moustakas}, {Bradley}, \& {Ford}}]{donahue15}
{Donahue}, M., {Connor}, T., {Fogarty}, K., {et~al.} 2015, \apj, 805, 177

\bibitem[{{Dopita} {et~al.}(2013){Dopita}, {Sutherland}, {Nicholls}, {Kewley},
  \& {Vogt}}]{dopita13}
{Dopita}, M.~A., {Sutherland}, R.~S., {Nicholls}, D.~C., {Kewley}, L.~J., \&
  {Vogt}, F.~P.~A. 2013, \apjs, 208, 10

\bibitem[{{Ebneter} \& {Balick}(1985)}]{ebneter85}
{Ebneter}, K. \& {Balick}, B. 1985, \aj, 90, 183

\bibitem[{{Ebneter} {et~al.}(1988){Ebneter}, {Djorgovski}, \&
  {Davis}}]{ebneter88}
{Ebneter}, K., {Djorgovski}, S., \& {Davis}, M. 1988, \aj, 95, 422

\bibitem[{{Ferland} {et~al.}(2016){Ferland}, {Henney}, {O'Dell}, \&
  {Peimbert}}]{ferland16}
{Ferland}, G.~J., {Henney}, W.~J., {O'Dell}, C.~R., \& {Peimbert}, M. 2016,
  \rmxaa, 52, 261

\bibitem[{{Ferrari} {et~al.}(1999){Ferrari}, {Pastoriza}, {Macchetto}, \&
  {Caon}}]{ferrari99}
{Ferrari}, F., {Pastoriza}, M.~G., {Macchetto}, F., \& {Caon}, N. 1999, \aaps,
  136, 269

\bibitem[{{Finkelman} {et~al.}(2012){Finkelman}, {Brosch}, {Funes}, {Barway},
  {Kniazev}, \& {V{\"a}is{\"a}nen}}]{finkelman12}
{Finkelman}, I., {Brosch}, N., {Funes}, J.~G., {et~al.} 2012, \mnras, 422, 1384

\bibitem[{{Fogarty} {et~al.}(2015){Fogarty}, {Postman}, {Connor}, {Donahue}, \&
  {Moustakas}}]{fogarty15}
{Fogarty}, K., {Postman}, M., {Connor}, T., {Donahue}, M., \& {Moustakas}, J.
  2015, \apj, 813, 117

\bibitem[{{Gaspari} {et~al.}(2013){Gaspari}, {Ruszkowski}, \& {Oh}}]{gaspari13}
{Gaspari}, M., {Ruszkowski}, M., \& {Oh}, S.~P. 2013, \mnras, 432, 3401

\bibitem[{{Gaspari} {et~al.}(2017){Gaspari}, {Temi}, \&
  {Brighenti}}]{gaspari17}
{Gaspari}, M., {Temi}, P., \& {Brighenti}, F. 2017, \mnras, 466, 677

\bibitem[{{Ginolfi} {et~al.}(2017){Ginolfi}, {Maiolino}, {Nagao}, {Carniani},
  {Belfiore}, {Cresci}, {Hatsukade}, {Mannucci}, {Marconi}, {Pallottini},
  {Schneider}, \& {Santini}}]{ginolfi17}
{Ginolfi}, M., {Maiolino}, R., {Nagao}, T., {et~al.} 2017, \mnras, 468, 3468

\bibitem[{{Goudfrooij} {et~al.}(1994){Goudfrooij}, {Hansen}, {Jorgensen}, \&
  {Norgaard-Nielsen}}]{goudfrooij94}
{Goudfrooij}, P., {Hansen}, L., {Jorgensen}, H.~E., \& {Norgaard-Nielsen},
  H.~U. 1994, \aaps, 105, 341

\bibitem[{{Graham} {et~al.}(2011){Graham}, {Onken}, {Athanassoula}, \&
  {Combes}}]{graham11}
{Graham}, A.~W., {Onken}, C.~A., {Athanassoula}, E., \& {Combes}, F. 2011,
  \mnras, 412, 2211

\bibitem[{{Grillmair} {et~al.}(1994){Grillmair}, {Faber}, {Lauer}, {Baum},
  {Lynds}, {O'Neil}, \& {Shaya}}]{grillmair94}
{Grillmair}, C.~J., {Faber}, S.~M., {Lauer}, T.~R., {et~al.} 1994, \aj, 108,
  102

\bibitem[{{Hayakawa} {et~al.}(2006){Hayakawa}, {Hoshino}, {Ishida}, {Furusho},
  {Yamasaki}, \& {Ohashi}}]{hayakawa06}
{Hayakawa}, A., {Hoshino}, A., {Ishida}, M., {et~al.} 2006, \pasj, 58, 695

\bibitem[{{Hempel} {et~al.}(2005){Hempel}, {Geisler}, {Hoard}, \&
  {Harris}}]{hempel05}
{Hempel}, M., {Geisler}, D., {Hoard}, D.~W., \& {Harris}, W.~E. 2005, \aap,
  439, 59

\bibitem[{{Hilker} {et~al.}(2018){Hilker}, {Richtler}, {Barbosa}, {Arnaboldi},
  {Coccato}, \& {Mendes de Oliveira}}]{hilker18}
{Hilker}, M., {Richtler}, T., {Barbosa}, C.~E., {et~al.} 2018, \aap, 619, A70

\bibitem[{{Hirashita} \& {Nozawa}(2017)}]{hirashita17}
{Hirashita}, H. \& {Nozawa}, T. 2017, \planss, 149, 45

\bibitem[{{Hirashita} {et~al.}(2015){Hirashita}, {Nozawa}, {Villaume}, \&
  {Srinivasan}}]{hirashita15}
{Hirashita}, H., {Nozawa}, T., {Villaume}, A., \& {Srinivasan}, S. 2015,
  \mnras, 454, 1620

\bibitem[{{H{\"o}nig} \& {Kishimoto}(2017)}]{hoenig17}
{H{\"o}nig}, S.~F. \& {Kishimoto}, M. 2017, \apjl, 838, L20

\bibitem[{{H{\"o}nig} {et~al.}(2013){H{\"o}nig}, {Kishimoto}, {Tristram},
  {Prieto}, {Gandhi}, {Asmus}, {Antonucci}, {Burtscher}, {Duschl}, \&
  {Weigelt}}]{hoenig13}
{H{\"o}nig}, S.~F., {Kishimoto}, M., {Tristram}, K.~R.~W., {et~al.} 2013, \apj,
  771, 87

\bibitem[{{Husemann} {et~al.}(2016){Husemann}, {Scharw{\"a}chter}, {Bennert},
  {Mainieri}, {Woo}, \& {Kakkad}}]{husemann16}
{Husemann}, B., {Scharw{\"a}chter}, J., {Bennert}, V.~N., {et~al.} 2016, \aap,
  594, A44

\bibitem[{{Jeong} {et~al.}(2013){Jeong}, {Yi}, {Kyeong}, {Sarzi}, {Sung}, \&
  {Oh}}]{jeong13}
{Jeong}, H., {Yi}, S.~K., {Kyeong}, J., {et~al.} 2013, \apjs, 208, 7

\bibitem[{{Kauffmann} {et~al.}(2003){Kauffmann}, {Heckman}, {Tremonti},
  {Brinchmann}, {Charlot}, {White}, {Ridgway}, {Brinkmann}, {Fukugita}, {Hall},
  {Ivezi{\'c}}, {Richards}, \& {Schneider}}]{kauffmann03}
{Kauffmann}, G., {Heckman}, T.~M., {Tremonti}, C., {et~al.} 2003, \mnras, 346,
  1055

\bibitem[{{Kewley} \& {Dopita}(2002)}]{kewley02}
{Kewley}, L.~J. \& {Dopita}, M.~A. 2002, \apjs, 142, 35

\bibitem[{{Kleiner} {et~al.}(2017){Kleiner}, {Pimbblet}, {Jones}, {Koribalski},
  \& {Serra}}]{kleiner17}
{Kleiner}, D., {Pimbblet}, K.~A., {Jones}, D.~H., {Koribalski}, B.~S., \&
  {Serra}, P. 2017, \mnras, 466, 4692

\bibitem[{{Konigl} \& {Kartje}(1994)}]{koenigl94}
{Konigl}, A. \& {Kartje}, J.~F. 1994, \apj, 434, 446

\bibitem[{{Kulkarni} {et~al.}(2014){Kulkarni}, {Sahu}, {Chaware},
  {Chakradhari}, \& {Pandey}}]{kulkarni14}
{Kulkarni}, S., {Sahu}, D.~K., {Chaware}, L., {Chakradhari}, N.~K., \&
  {Pandey}, S.~K. 2014, \na, 30, 51

\bibitem[{{Lauer} {et~al.}(2005){Lauer}, {Faber}, {Gebhardt}, {Richstone},
  {Tremaine}, {Ajhar}, {Aller}, {Bender}, {Dressler}, {Filippenko}, {Green},
  {Grillmair}, {Ho}, {Kormendy}, {Magorrian}, {Pinkney}, \& {Siopis}}]{lauer05}
{Lauer}, T.~R., {Faber}, S.~M., {Gebhardt}, K., {et~al.} 2005, \aj, 129, 2138

\bibitem[{{Lindblad}(1977)}]{lindblad77}
{Lindblad}, P.~O. 1977, The Messenger, 10, 20

\bibitem[{{Macchetto} {et~al.}(1996){Macchetto}, {Pastoriza}, {Caon}, {Sparks},
  {Giavalisco}, {Bender}, \& {Capaccioli}}]{macchetto96}
{Macchetto}, F., {Pastoriza}, M., {Caon}, N., {et~al.} 1996, \aaps, 120, 463

\bibitem[{{Marigo} {et~al.}(2017){Marigo}, {Girardi}, {Bressan}, {Rosenfield},
  {Aringer}, {Chen}, {Dussin}, {Nanni}, {Pastorelli}, {Rodrigues}, {Trabucchi},
  {Bladh}, {Dalcanton}, {Groenewegen}, {Montalb{\'a}n}, \& {Wood}}]{marigo17}
{Marigo}, P., {Girardi}, L., {Bressan}, A., {et~al.} 2017, \apj, 835, 77

\bibitem[{{McDonald} {et~al.}(2015){McDonald}, {McNamara}, {van Weeren},
  {Applegate}, {Bayliss}, {Bautz}, {Benson}, {Carlstrom}, {Bleem}, {Chatzikos},
  {Edge}, {Fabian}, {Garmire}, {Hlavacek-Larrondo}, {Jones-Forman}, {Mantz},
  {Miller}, {Stalder}, {Veilleux}, \& {ZuHone}}]{mcdonald15}
{McDonald}, M., {McNamara}, B.~R., {van Weeren}, R.~J., {et~al.} 2015, \apj,
  811, 111

\bibitem[{{Micha{\l}owski} {et~al.}(2019){Micha{\l}owski}, {Hjorth}, {Gall},
  {Frayer}, {Tsai}, {Hirashita}, {Rowland s}, {Takeuchi}, {Le{\'s}niewska},
  {Behrendt}, {Bourne}, {Hughes}, {Spring}, {Zavala}, \&
  {Bartczak}}]{michalowski19}
{Micha{\l}owski}, M.~J., {Hjorth}, J., {Gall}, C., {et~al.} 2019, arXiv
  e-prints, arXiv:1910.06327

\bibitem[{{Misgeld} {et~al.}(2011){Misgeld}, {Mieske}, {Hilker}, {Richtler},
  {Georgiev}, \& {Schuberth}}]{misgeld11}
{Misgeld}, I., {Mieske}, S., {Hilker}, M., {et~al.} 2011, \aap, 531, A4

\bibitem[{{Momcheva} {et~al.}(2013){Momcheva}, {Lee}, {Ly}, {Salim}, {Dale},
  {Ouchi}, {Finn}, \& {Ono}}]{momcheva13}
{Momcheva}, I.~G., {Lee}, J.~C., {Ly}, C., {et~al.} 2013, \aj, 145, 47

\bibitem[{{Mor} {et~al.}(2009){Mor}, {Netzer}, \& {Elitzur}}]{mor09}
{Mor}, R., {Netzer}, H., \& {Elitzur}, M. 2009, \apj, 705, 298

\bibitem[{{Murga} {et~al.}(2015){Murga}, {Zhu}, {M{\'e}nard}, \&
  {Lan}}]{murga15}
{Murga}, M., {Zhu}, G., {M{\'e}nard}, B., \& {Lan}, T.-W. 2015, \mnras, 452,
  511

\bibitem[{{Olivares} {et~al.}(2019){Olivares}, {Salome}, {Combes}, {Hamer},
  {Guillard}, {Lehnert}, {Polles}, {Beckmann}, {Dubois}, {Donahue}, {Edge},
  {Fabian}, {McNamara}, {Rose}, {Russell}, {Tremblay}, {Vantyghem}, {Canning},
  {Ferland }, {Godard}, {Peirani}, \& {Pineau des Forets}}]{olivares19}
{Olivares}, V., {Salome}, P., {Combes}, F., {et~al.} 2019, \aap, 631, A22

\bibitem[{{Park} {et~al.}(2015){Park}, {Jeong}, \& {Yi}}]{park15}
{Park}, J., {Jeong}, H., \& {Yi}, S.~K. 2015, \apj, 809, 91

\bibitem[{{Proxauf} {et~al.}(2014){Proxauf}, {{\"O}ttl}, \&
  {Kimeswenger}}]{proxauf14}
{Proxauf}, B., {{\"O}ttl}, S., \& {Kimeswenger}, S. 2014, \aap, 561, A10

\bibitem[{{Prugniel} \& {Heraudeau}(1998)}]{prugniel98}
{Prugniel}, P. \& {Heraudeau}, P. 1998, \aaps, 128, 299

\bibitem[{{Richtler} {et~al.}(2011){Richtler}, {Salinas}, {Misgeld}, {Hilker},
  {Hau}, {Romanowsky}, {Schuberth}, \& {Spolaor}}]{richtler11}
{Richtler}, T., {Salinas}, R., {Misgeld}, I., {et~al.} 2011, \aap, 531, A119+

\bibitem[{{Rieke} \& {Lebofsky}(1985)}]{rieke85}
{Rieke}, G.~H. \& {Lebofsky}, M.~J. 1985, \apj, 288, 618

\bibitem[{{Runge} \& {Yan}(2018)}]{runge18}
{Runge}, J. \& {Yan}, H. 2018, \apj, 853, 47

\bibitem[{{Sadler} \& {Gerhard}(1985)}]{sadler85}
{Sadler}, E.~M. \& {Gerhard}, O.~E. 1985, \mnras, 214, 177

\bibitem[{{Sancisi} {et~al.}(2008){Sancisi}, {Fraternali}, {Oosterloo}, \& {van
  der Hulst}}]{sancisi08}
{Sancisi}, R., {Fraternali}, F., {Oosterloo}, T., \& {van der Hulst}, T. 2008,
  \aapr, 15, 189

\bibitem[{{Sansom} {et~al.}(2019){Sansom}, {Glass}, {Bendo}, {Davis},
  {Rowlands}, {Bourne}, {Dunne}, {Eales}, {Kaviraj}, {Popescu}, {Smith}, \&
  {Viaene}}]{sansom19}
{Sansom}, A.~E., {Glass}, D.~H.~W., {Bendo}, G.~J., {et~al.} 2019, \mnras, 482,
  4617

\bibitem[{{Schultz} \& {Wiemer}(1975)}]{schultz75}
{Schultz}, G.~V. \& {Wiemer}, W. 1975, \aap, 43, 133

\bibitem[{{Smith} \& {Edge}(2017)}]{smith17}
{Smith}, R.~J. \& {Edge}, A.~C. 2017, \mnras, 471, L66

\bibitem[{{Sparks} {et~al.}(1997){Sparks}, {Marcella Carollo}, \&
  {Macchetto}}]{sparks97}
{Sparks}, W.~B., {Marcella Carollo}, C., \& {Macchetto}, F.~o. 1997, \apj, 486,
  253

\bibitem[{{Tran} {et~al.}(2001){Tran}, {Tsvetanov}, {Ford}, {Davies}, {Jaffe},
  {van den Bosch}, \& {Rest}}]{tran01}
{Tran}, H.~D., {Tsvetanov}, Z., {Ford}, H.~C., {et~al.} 2001, \aj, 121, 2928

\bibitem[{{van Dokkum} \& {Franx}(1995)}]{vandokkum95}
{van Dokkum}, P.~G. \& {Franx}, M. 1995, \aj, 110, 2027

\bibitem[{{Vasterberg} {et~al.}(1991){Vasterberg}, {Jorsater}, \&
  {Lindblad}}]{vasterberg91}
{Vasterberg}, A.~R., {Jorsater}, S., \& {Lindblad}, P.~O. 1991, \aap, 247, 335

\bibitem[{{Ventimiglia} {et~al.}(2011){Ventimiglia}, {Arnaboldi}, \&
  {Gerhard}}]{ventimiglia11}
{Ventimiglia}, G., {Arnaboldi}, M., \& {Gerhard}, O. 2011, \aap, 528, A24

\bibitem[{{Ventimiglia} {et~al.}(2010){Ventimiglia}, {Gerhard}, {Arnaboldi}, \&
  {Coccato}}]{ventimiglia10}
{Ventimiglia}, G., {Gerhard}, O., {Arnaboldi}, M., \& {Coccato}, L. 2010, \aap,
  520, L9

\bibitem[{{Vogelsberger} {et~al.}(2019){Vogelsberger}, {McKinnon}, {O'Neil},
  {Marinacci}, {Torrey}, \& {Kannan}}]{vogelsberger19}
{Vogelsberger}, M., {McKinnon}, R., {O'Neil}, S., {et~al.} 2019, \mnras, 487,
  4870

\bibitem[{{Walcher} {et~al.}(2015){Walcher}, {Coelho}, {Gallazzi}, {Bruzual},
  {Charlot}, \& {Chiappini}}]{walcher15}
{Walcher}, C.~J., {Coelho}, P.~R.~T., {Gallazzi}, A., {et~al.} 2015, \aap, 582,
  A46

\bibitem[{{Wirth} \& {Gallagher}(1980)}]{wirth80}
{Wirth}, A. \& {Gallagher}, J.~S. 1980, \apj, 242, 469

\end{thebibliography}
  
\end{document}